\documentclass[aps,prl,twocolumn,superscriptaddress]{revtex4-2}

\providecommand{\bibcommenthead}[1]{}

\usepackage{amsmath}
\usepackage{color}

\usepackage{graphicx}
\usepackage[inline,shortlabels]{enumitem}
\usepackage{soul}

\usepackage[colorlinks,
linkcolor=black,
anchorcolor=black,
urlcolor=black,
citecolor=black]{hyperref}

\begin{document}

\title{A mechanical approach to facilitate the formation of dodecagonal quasicrystals and their approximants}

\author{Zhehua Jiang}
\affiliation{Hefei National Research Center for Physical Sciences at the Microscale and CAS Key Laboratory of Microscale Magnetic Resonance, University of Science and Technology of China, Hefei 230026, People’s Republic of China}
\affiliation{Department of Physics, University of Science and Technology of China, Hefei 230026, People’s Republic of China}

\author{Jianhua Zhang}
\affiliation{Department of Physics, University of Science and Technology of China, Hefei 230026, People’s Republic of China}

\author{Mengyuan Zhan}
\author{Jiaqi Si}
\author{Junchao Huang}
\author{Hua Tong}
\email[Email:]{huatong@ustc.edu.cn}
\author{Ning Xu}
\email[Email:]{ningxu@ustc.edu.cn}
\affiliation{Hefei National Research Center for Physical Sciences at the Microscale and CAS Key Laboratory of Microscale Magnetic Resonance, University of Science and Technology of China, Hefei 230026, People’s Republic of China}
\affiliation{Department of Physics, University of Science and Technology of China, Hefei 230026, People’s Republic of China}

\begin{abstract}
The conditions for forming quasicrystals and their approximants are stringent, normally requiring multiple length scales to stabilize the quasicrystalline order. Here we report an unexpected finding that the approximants and motifs of dodecagonal quasicrystals can be spontaneously formed in the simplest system of identical hard disks, utilizing the unstable feature of the initial square packing subject to mechanical perturbations. Because there is only one length scale involved, this finding challenges existing theories of quasicrystals and their approximants. By applying the same approach to a system known to form a dodecagonal quasicrystal, we develop decent quasicrystalline order in a purely mechanical manner. With the aid of thermal treatment, we achieve a significantly better quasicrystalline order than that from the direct self-assembly of the liquid state within the same period of time. In sufficiently low temperatures where the self-assembly of a liquid is significantly hindered, our approach still promotes the formation of quasicrystals. Our study thus opens a venue for high-efficiency search and formation of quasicrystals, and may have broader implications for the design and synthesis of quasicrystalline materials.      
\end{abstract}


\maketitle


Quasicrystals, first discovered by Shechtman \textit{et al.} \cite{Shechtman_Blech_Gratias_Cahn_1984}, exhibit sharp Bragg peaks with symmetries forbidden in ordinary crystals and lack long-range translational order \cite{Levine_Steinhardt_1984}. For the first two decades, quasicrystalline orders were mainly observed in metallic alloys, until the discovery of quasicrystals in soft matter in 2004 \cite{Zeng_Ungar_Liu_Percec_Dulcey_Hobbs_2004}. The building blocks of soft matter may span multiple scales, making it a playground for diverse physical phenomena \cite{Saarloos_Vitelli_Zeravcic_2024} and providing vast possibilities to tune interparticle interactions and the shape of structural units for the formation of quasicrystals \cite{Ungar_Zeng_2005,Dotera_2011,Su_Zhang_Yan_Guo_Huang_Shan_Liu_Liu_Huang_Cheng_2020,Nagaoka_Schneider_Zhu_Chen_2023}. Consequently, soft quasicrystals \cite{Lifshitz_Diamant_2007} have been observed in a wide range of soft matter systems, including dendrimer liquid crystals \cite{Zeng_Ungar_Liu_Percec_Dulcey_Hobbs_2004}, star-shaped polymers \cite{Hayashida_Dotera_Takano_Matsushita_2007}, binary nanoparticle systems \cite{Talapin_Shevchenko_Bodnarchuk_Ye_Chen_Murray_2009}, block copolymer micelles \cite{Fischer_Exner_Zielske_Perlich_Deloudi_Steurer_Lindner_Förster_2011}, mesoporous silica \cite{Xiao_Fujita_Miyasaka_Sakamoto_Terasaki_2012}, DNA-engineered biomolecules \cite{Zhou_Lim_Lin_Lee_Li_Huang_Du_Lee_Wang_Sánchez-Iglesias_et_al._2024} and microspheres in applied magnetic and electric fields \cite{Gao_Sprinkle_Marr_Wu_2025}. 

Compared to quasicrystals, their approximants, which share the same prototiles but exhibit long-range translational order, are easier to form. This makes quasicrystal approximants an ideal starting point for elucidating both the formation mechanism and  structural attributes of quasicrystals \cite{Chen_Li_Kuo_1988,Keys_Glotzer_2007,Hayashida_Dotera_Takano_Matsushita_2007,Lee_Bluemle_Bates_2010,Schenk_Krahn_Cockayne_Meyerheim_de_Boissieu_Förster_Widdra_2022,Mueller_Lindsay_Lewis_Zhang_Narayanan_Lodge_Mahanthappa_Bates_2024}. Owing to their unique structural and physical properties, such as exceptional hardness, low friction, and thermal stability, quasicrystals and their approximants hold promising applications in thermal insulation, surface coatings, and advanced materials engineering \cite{Yadav_Mukhopadhyay_2018}.  

Although the structures of quasicrystals and their approximants have been extensively studied \cite{Walter_Deloudi_2009}, their formation mechanisms are not yet fully understood \cite{Steurer_2018,Je_Lee_Teich_Engel_Glotzer_2021, Yin_Jiang_Shi_Zhang_Zhang_2021}. Typically, multiple characteristic length scales are involved in the structures of quasicrystals and their approximants. This leads to the prevailing hypothesis that their formation necessitates the introduction of multiple length scales. Consequently, studies often employ complex interparticle interactions with multiple length scales \cite{Lifshitz_Diamant_2007, Dotera_2011, Engel_Trebin_2007, Dotera_Oshiro_Ziherl_2014, Barkan_Engel_Lifshitz_2014, Engel_Damasceno_Phillips_Glotzer_2015, Schoberth_Emmerich_Holzinger_Dulle_Förster_Gruhn_2016}, polydisperse isotropic particles  \cite{Fayen_Impéror-Clerc_Filion_Foffi_Smallenburg_2023, Bedolla-Montiel_Lange_Ortiz_Dijkstra_2024}, or anisotropic particles with patches \cite{Reinhardt_Romano_Doye_2013,Noya_Wong_Llombart_Doye_2021} or polyhedral shapes \cite{Haji-Akbari_Engel_Keys_Zheng_Petschek_Palffy-Muhoray_Glotzer_2009,Je_Lee_Teich_Engel_Glotzer_2021}. 

A notable exception to this general view was reported recently: both dodecagonal and octagonal quasicrystals can be formed by monodisperse isotropic disks interacting via simple spring-like repulsions \cite{Zu_Tan_Xu_2017,Fomin_Gaiduk_Tsiok_Ryzhov_2018}. In this case, quasicrystals are formed without explicitly engineering multiple length scales, but instead relies on the high density of the system: the soft-core nature of the interactions promotes the self-assembly of disks into pentagons, which then serve as the fundamental building blocks for quasicrystal formation. 

Note that disks interacting via spring-like repulsions effectively behave as hard ones at sufficiently low pressures \cite{Xu_Haxton_Liu_Nagel_2009,Xu_2019}. Here we address a more challenging question: Can the simplest system of monodisperse hard disks spontaneously form quasicrystals or their approximants? Given the presence of only one length scale$-$the hard disk diameter$-$this question may sound counterintuitive based on previous understanding. It is an indisputable fact that the thermodynamic equilibrium state of monodisperse hard disks is the hexagonal phase \cite{Royall_Charbonneau_Dijkstra_Russo_Smallenburg_Speck_Valeriani_2024}. Although other metastable states, such as polycrystalline packings of hard disks, can be obtained via rapid compression from liquid states, to our knowledge, no spontaneous formation of complex crystalline or quasicrystalline structures has ever been reported, even under nonequilibrium conditions.

\begin{figure*}
    \centering
    \includegraphics[width=0.95\linewidth]{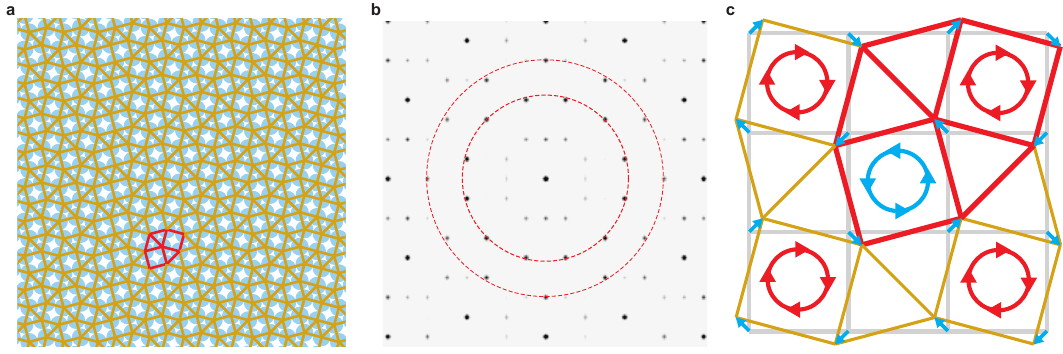}
    \caption{{\bf Formation of a DDQC approximant using hard disks and the transformation pathway.} {\bf a} DDQC approximant obtained from the perturbation of a square packing. Circles and lines represent disks and bonds, respectively. Red bonds highlight the $(3^2.4.3.4)$ Archimedean tiling. {\bf b} Diffraction pattern of the approximant shown in {\bf a}. While 12-fold-like symmetry is visually apparent, close inspection reveals only four diffraction spots lying on each ring (red dashed circles), indicating a 4-fold symmetry. {\bf c} Local transformation pathway from a square packing to the approximant. Gray bonds represent the initial square packing. Orange and red bonds represent the approximant.  Arrows indicate the displacements of disks over the transformation. Directed circles illustrate the rotational direction of the squares. Red bonds highlight the $(3^2.4.3.4)$ Archimedean tiling.  }
    \label{fig:fig1}
\end{figure*}

However, a straightforward analysis suggests that the idea might not be totally infeasible. Stampfli-G{\"a}hler tiling of a dodecagonal quasicrystal requires only two building blocks: squares and equilateral triangles \cite{Stampfli1986ADQ, gahler1988quasicrystalline}. A hexagonal packing of hard disks is composed of equilateral triangles. Hard disks can also be arranged into a square lattice, which is however unstable due to the presence of soft modes, i.e., normal modes with zero frequency or energy. Therefore, these two building blocks can be constructed using hard disks, albeit at different packing densities. The critical challenge is to determine how to make these two building blocks spontaneously coexist and assemble into more complex motifs of quasicrystals or their approximants.

Utilizing the unstable feature of the square lattice, we unexpectedly observe the spontaneous formation of the approximant of the dodecagonal quasicrystal (DDQC) via a slight perturbation of the square packing of hard disks at constant pressure. By selectively removing pairs of disks from the square lattice, the same protocol results in the spontaneous formation of DDQC motifs.  Although global formation of DDQCs  has not yet been achieved with hard disks, this protocol significantly accelerates the formation of DDQCs in known DDQC-forming systems. Our study reveals the previously unrecognized capability of hard disks to assemble into complex crystals and identifies a novel pathway of the formation of DDQCs and their approximants, providing a highly efficient and mechanical method to facilitate the formation and search for DDQCs.    

\begin{figure*}
    \centering
    \includegraphics[width=0.95\linewidth]{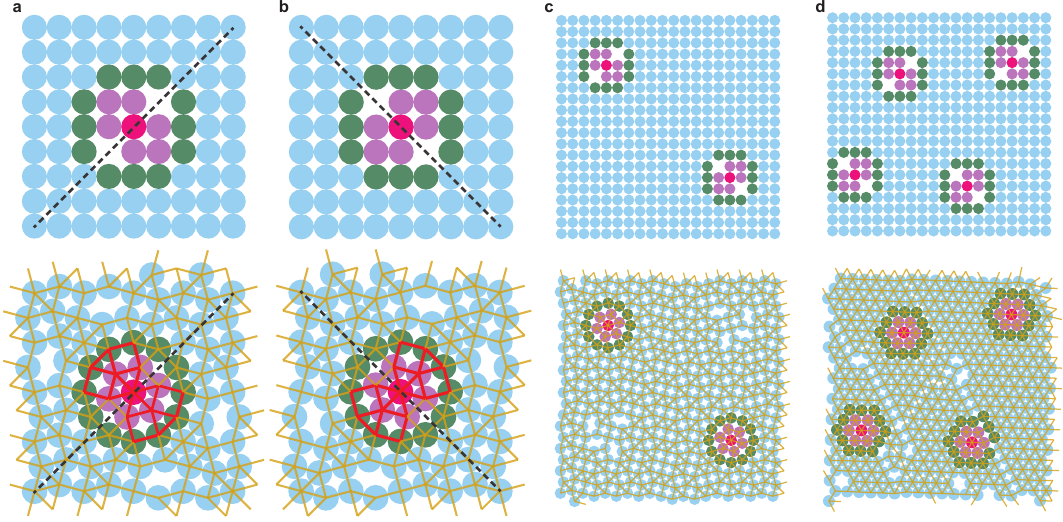}
    \caption{{\bf Formation of DDQC motifs using hard disks.} {\bf a} Introduction of a vacancy pair along the $45^{\circ}$ direction (dashed line) to a square packing of $N=81$ disks (top panel) and the resulting DDQC motif (bottom panel). Disks transformed to the center, first shell, and second shell of the motif are shown in red, purple, and green, respectively. Lines in the bottom panel represent bonds, with red bonds highlighting two head-to-head $(3^2.4.3.4)$ Archimedean tilings. {\bf b} Generation of another DDQC motif with a $30^{\circ}$ rotation relative to that in {\bf a}, achieved by removing a vacancy pair along the $-45^{\circ}$ direction (dashed line). {\bf c,d} Introduction of $2$ and $4$ vacancy pairs, respectively, to an $N=400$ square packing (top panel) and the packing after the transformation (bottom panel).    
    }
    \label{fig:fig2}
\end{figure*}

\vspace{2mm}
\noindent {\bf \large{Results}}
\vspace{1mm}

\noindent {\bf Forming DDQC approximant by hard disks}

\noindent We first study static packings of disks interacting via the harmonic repulsion (detailed in Methods). At low pressures, these packings approximate hard disk systems, as the harmonic repulsion effectively mimics the hard-core exclusion. We begin with a square packing of disks at a pressure of $p=10^{-4}$. The square packing is unstable subject to any infinitesimal perturbations due to the presence of soft modes. Here we introduce the perturbation by randomly displacing a disk slightly away from its equilibrium position. We have confirmed that other types of small perturbations yield equivalent results. After minimizing the enthalpy (detailed in Methods) at fixed pressure, we obtain a mechanically stable packing shown in Fig.~\ref{fig:fig1}a. By connecting contacting particles with bonds, the mixing tiling of squares and equilateral triangles can be clearly observed, demonstrating the spontaneous formation of a complex ordered structure.  

As highlighted by the red bonds in Fig.~\ref{fig:fig1}a, a fundamental Archimedean tiling motif in our packing is ($3^2.4.3.4$), a prototile also prevalent in DDQCs. Fig.~\ref{fig:fig1}b shows the diffraction pattern of the packing. Although it exhibits a 12-fold-like arrangement, as illustrated by the red dashed rings, only four diffraction spots lie on each ring, suggesting a deviation from perfect 12-fold symmetry. Therefore, the packing is identified as a DDQC approximant with 4-fold symmetry. The periodicity of the structure can be observed in Fig.~\ref{fig:fig1}a, and the number ratio of triangles to squares is $2$, differing from $4/\sqrt{3}$ for perfect DDQCs \cite{Walter_Deloudi_2009,Oxborrow_Henley_1993}.  All these characteristics indicate a crystalline rather than quasicrystalline order. Nevertheless, it is rather unexpected that simplest monodisperse hard disks can spontaneously form such complex crystals. 

To illustrate the transformation from the square packing to the DDQC approximant, we show the displacement field in Fig.~\ref{fig:fig1}c. We focus on a local $4\times 4$ lattice region which includes $9$ adjacent squares. The arrows represent disk movements, with the tail and head indicating the initial and final positions, respectively. The central square rotates either clockwise or counter-clockwise, while the four corner squares rotate in the opposite direction, creating a coordinated local rearrangement. Consequently, the four squares adjacent to the central square are separately squeezed into a pair of edge-sharing triangles. 

\begin{figure*}
    \centering
    \includegraphics[width=1\linewidth]{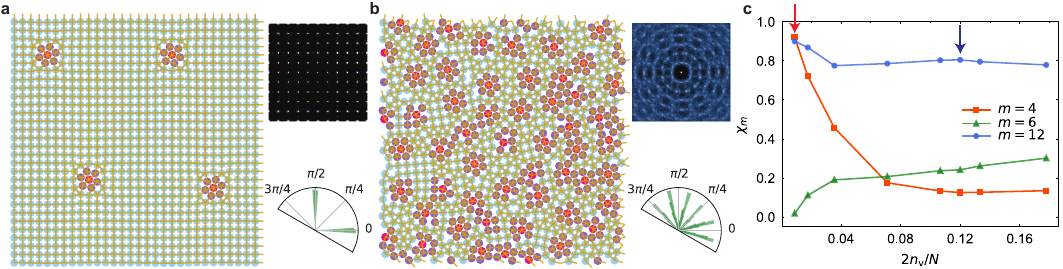}
    \caption{{\bf Formation of DDQCs using TLS disks through a mechanical approach.} {\bf a,b} Results for introducing $n_{\rm v}=4$ and $54$ vacancy pairs, respectively, to an $N=900$ square packing, followed by energy minimization. The left panel shows the configuration with bonds. Disks in red and purple are those expected to become the center and first shell of DDQC motifs. The right top and right bottom panels show the diffraction pattern and the distribution of bond angles, respectively. {\bf c} Evolution of the $m$-fold orientational order $\chi_m$ for the resultant packings with $2n_{\rm v}/ N$. The two arrows point to $n_{\rm v}=4$ and $54$ as shown in {\bf a} and {\bf b}, respectively. }
    \label{fig:fig3}
\end{figure*}

\vspace{2mm}

\noindent {\bf Forming DDQC motifs by hard disks}

\noindent The spontaneous formation of the DDQC approximant suggests an intriguing possibility of forming DDQCs using hard disks. We note that the fundamental motif of Stampfli-G{\"a}hler tiling of DDQCs is composed of $19$ disks arranged in a specific geometry, as illustrated in the bottom panel of Fig.~\ref{fig:fig2}a. A central disk (red) is surrounded by two concentric polygons: an inner hexagon (purple) and an outer dodecagon (green). The square-triangle tessellation of this motif includes two head-to-head ($3^2.4.3.4$) Archimedean tilings. Based on the mechanism depicted in Fig.~\ref{fig:fig1}c, the formation of this pattern necessarily requires a local displacement field with mirror symmetry. 

Figure~\ref{fig:fig2}a illustrates our strategy to induce particle motion with mirror symmetry. As shown in the top panel, we remove a pair of disks from the square packing, which are the second-nearest neighbors of a central disk (red) and align along the $45^{\circ}$ dashed line passing the central disk. This removal disrupts the force balance on the eight disks immediately surrounding these two vacancies just created, leading to particle rearrangements under constant pressure. Apparently, the presence of these two vacancies will induce mirror-symmetric movements of disks about the $45^{\circ}$ line. Interestingly, following the enthalpy minimization, a DDQC motif is successfully produced, as shown in the bottom panel of Fig.~\ref{fig:fig2}a. Surrounding the motif, various tiling patterns emerge, including the ($3^2.4.3.4$) tiling. Figure~\ref{fig:fig2}b demonstrates that removing a pair of disks along the $-45^{\circ}$ line results in another motif with a $30^{\circ}$ rotation relative to that in Fig.~\ref{fig:fig2}a. Both motifs are characteristic of the Stampfli-G{\"a}hler tiling of perfect DDQCs. Since each motif alone produces only 6-fold dihedral symmetry, their combination is essential to form the 12-fold symmetry in perfect DDQCs \cite{Ishimasa_Iwami_Sakaguchi_Oota_Mihalkovič_2015,Impéror-Clerc_Jagannathan_Kalugin_Sadoc_2021}. Therefore, our mechanical approach effectively generates both motifs, providing a potential pathway to construct DDQC structures.   

The question now becomes: can a long-range DDQC order be realized by introducing a sufficient number of vacancy pairs into the square packing? We start the investigation by randomly introducing $n_{\rm v}$ pairs of both types of vacancies into the system, and distributing them uniformly in space.  As shown in Fig.~\ref{fig:fig2}c, for a square packing of $N=400$ hard disks, two DDQC motifs are generated when $n_{\rm v}=2$. However, when $n_{\rm v}= 4$, Fig.~\ref{fig:fig2}d indicates that our approach fails to generate or sustain DDQC motifs. Instead, the hexagonal packing dominates after the transformation. As shown in the Supplementary Movie, during the early stage of the transformation, ($3^2.4.3.4$) tilings emerge and grow, but are eventually overwhelmed by the rapid expansion of the hexagonal packing. For hard disks, the hexagonal packing is thermodynamically stable. In contrast, DDQC approximants and structures with DDQC motifs are metastable, being much less stable than the hexagonal packing and susceptible to destabilization under large perturbations. Increasing the number of vacancy pairs enhances the perturbation, destabilizing these complex structures and driving the system toward the more stable hexagonal phase. Therefore, the formation of quasicrystals requires more stringent conditions. If metastable DDQCs exist in hard disk systems, more refined protocols are required to control the local emergence and growth of quasicrystalline order, as well as the cooperative rearrangements in other regions.    

\begin{figure*}
    \centering
    \includegraphics[width=0.95\linewidth]{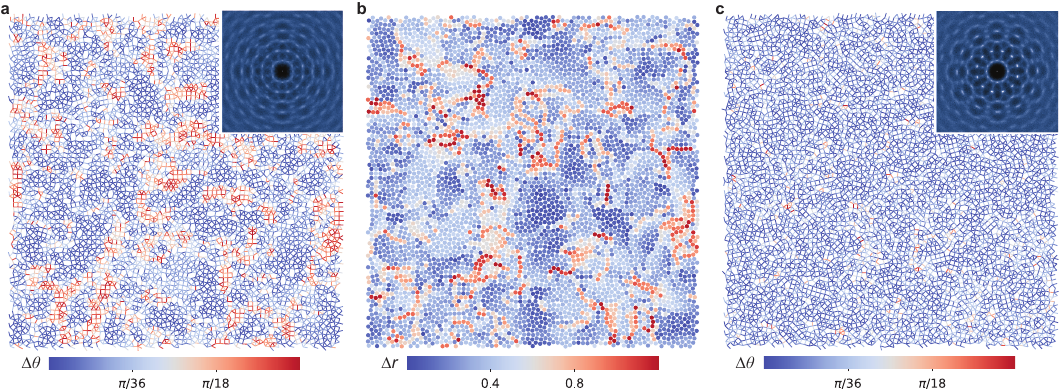}
    \caption{{\bf Enhancement of the DDQC order through the thermal treatment for systems with TLS disks.} {\bf a} Spatial distribution of the deviation of bond angle from the characteristic DDQC angles, $\Delta \theta$, for a packing mechanically transformed from an $N=6400$ square packing with $n_{\rm v}=427$ vacancy pairs. The color bar indicates the magnitude of $\Delta\theta$. {\bf b,c} Displacement field and spatial distribution of $\Delta\theta$, respectively, after equilibrating the state in {\bf a} over a time duration of $t=5\times10^{5}$ at $T=0.12$. The color bars in {\bf b} and {\bf c} indicate the magnitudes of particle displacement, $\Delta r$, and $\Delta\theta$, respectively. Diffraction patterns inserted in {\bf a} and {\bf c} highlight the growth of the DDQC order after the thermal treatment.}
    \label{fig:fig4}
\end{figure*}

\vspace{2mm}

\noindent {\bf Athermal growth of DDQC order in a DDQC-forming system}

\noindent Although our current mechanical approach fails to generate global DDQC order using hard disks, the successful generation of DDQC motifs suggests a potential avenue for DDQC exploration. To test this feasibility, we apply the same approach to a system with a two-length-scale (TLS) interaction potential (detailed in Methods). This system has been shown to self-assemble into a DDQC when equilibrated at appropriate densities and low temperatures \cite{Kryuchkov_Yurchenko_Fomin_Tsiok_Ryzhov_2018,Padilla_Ramírez-Hernández_2020,Coli_Boattini_Filion_Dijkstra_2022}. We initialize the system by arranging TLS disks into a square lattice and randomly introducing $n_{\rm v}$ vacancy pairs. We then adjust the number density to $\rho=0.94$, at which DDQC self-assembly can occur. An energy minimization is then performed to obtain static packings at $T=0$.    

As shown in the left panel of Fig.~\ref{fig:fig3}a, $n_{\rm v}=4$ vacancy pairs lead to four DDQC motifs in the static packing with initially $N=900$ TLS disks. Unlike hard disk systems, the square-lattice structure is still maintained around the motifs. This preservation of square-lattice order is attributed to the long-range particle interactions. Each disk interacts not only with its nearest neighbors but also with distant neighbors, stabilizing the square lattice structure. As a result, the 4-fold symmetry still dominates in the diffraction pattern, as confirmed in the right top panel of Fig.~\ref{fig:fig3}a. The right bottom panel of Fig.~\ref{fig:fig3}a illustrates the distribution of the bond angle $\theta$, with $\theta=0$ defined as the right horizontal direction. The bonds are predominantly aligned along the two principal directions of the square lattice, $\theta=0$ and $\pi/2$. As defined in Fig.~\ref{fig:fig2}a and \ref{fig:fig2}b, there are two types of DDQC motifs, contributing six characteristic DDQC angles: $-\pi/12$, $\pi/12$, $\pi/4$, $5\pi/12$, $7\pi/12$, and $3\pi/4$ \cite{gahler1988quasicrystalline,Walter_Deloudi_2009,Impéror-Clerc_Jagannathan_Kalugin_Sadoc_2021}. Since there are only four motifs, their contribution to the bond angle distribution is negligible in Fig.~\ref{fig:fig3}a.

Figure~\ref{fig:fig3}b shows the results for $n_{\rm v}=54$ vacancy pairs. Unlike the case of hard disks, the left panel illustrates that a significant fraction of DDQC motifs persists even at such a high concentration of vacancy pairs. The quasicrystalline order is evident in the diffraction pattern. Furthermore, the bond angle distribution reveals well-defined peaks corresponding to the six characteristic DDQC angles. Thus, by simply introducing vacancy pairs to a square lattice, the DDQC order is rapidly induced via an athermal approach.  

Figure~\ref{fig:fig3}c illustrates the evolution of the average local structural orders with $2n_{\rm v}/N$ for $N=900$. Here we present the results for 4-, 6-, and 12-fold symmetries, denoted as $\chi_4$, $\chi_6$, and $\chi_{12}$, respectively (detailed in Methods). For the initial square lattice ($n_{\rm v}=0$), both $\chi_4$ and $\chi_{12}$ equal $1$, while $\chi_6$ is almost zero, reflecting the incompatibility of the 6-fold symmetry with the 4-fold symmetry of square lattice. As $n_{\rm v}$ increases, both $\chi_4$ and $\chi_{12}$ decrease and saturate to a plateau. However, $\chi_4$ decreases to a low value of approximately $0.1$, whereas $\chi_{12}$ saturates at a high value of approximately $0.8$, indicating the breaking of the 4-fold symmetry and the emergence of the 12-fold symmetry. Meanwhile, $\chi_6$ gradually increases. Since hexagonal and square structures are fundamental elements in DDQCs, $\chi_4$ and $\chi_6$ remain nonzero, but are relatively small in DDQCs. For the $n_{\rm v}=54$ case shown in Fig.~\ref{fig:fig3}b, $\chi_{12}\approx 0.80$, $\chi_4\approx 0.12$, and $\chi_6\approx 0.26$, demonstrating that the observed 12-fold symmetry is primarily due to the DDQC order rather than the mixing contributions from 4- and 6-fold symmetries.     

\vspace{2mm}

\noindent {\bf Enhancing DDQC order via thermal treatment}

\noindent To quantify the quality of the DDQC order, we use the following metric. For a bond with angle $\theta$, we identify the closest of the six characteristic DDQC angles next to it, denoted as $\theta_{\rm c}$, and calculate the deviation $\Delta\theta=|\theta -\theta_{\rm c}|$. Therefore, smaller values of $\Delta\theta$ indicate better DDQC orders.  

Figure~\ref{fig:fig4}a presents a TLS disk packing obtained via our mechanical, athermal approach, where the DDQC order is evident from the diffraction pattern. This packing is induced by introducing $n_{\rm v}=427$ vacancy pairs to an $N=6400$ square packing. The bond color reflects its $\Delta \theta$ value. In regions where vacancy pairs are introduced and DDQC motifs are formed, the bonds generally have small values of $\Delta \theta$, whereas other regions show relatively larger values of $\Delta\theta$. Therefore, while the mechanical approach successfully induces the DDQC order, it is not sufficient to achieve optimal global order, indicating the necessity for additional treatments.   

We thus heat the packing to a low temperature of $T=0.12$ and equilibrate it for a duration of time $t$. Figure~\ref{fig:fig4}b illustrates the spatial distribution of the particle displacement over a duration of $t=5\times10^5$, with the color coding indicating the magnitude of particle displacement, $\Delta r$. A strong spatial correlation between $\Delta r$ and $\Delta \theta$ can be observed by comparing Fig.~\ref{fig:fig4}a and \ref{fig:fig4}b, with larger $\Delta r$ typically occurring in regions with relatively larger $\Delta \theta$. As illustrated in Fig.~\ref{fig:fig4}c, after equilibration, the majority of bonds exhibit small $\Delta \theta$ values, and the spatial distribution of $\Delta \theta$ becomes significantly more uniform. The optimization effect of the thermal treatment is further demonstrated by the sharpening of  spots in the diffraction pattern. 

\vspace{2mm}

\noindent {\bf Efficiency of the mechanical approach}

\noindent A notable advantage of the mechanical approach is the instantenous development of the DDQC order, significantly accelerating the DDQC formation. As shown above, the thermal treatment primarily serves to optimize regions with weak DDQC orders. It is thus expected that the overall process would be much less time-consuming than the direct formation of DDQCs from the liquid state at a specific temperature. 

\begin{figure}
    \centering
    \includegraphics[width=.95\linewidth]{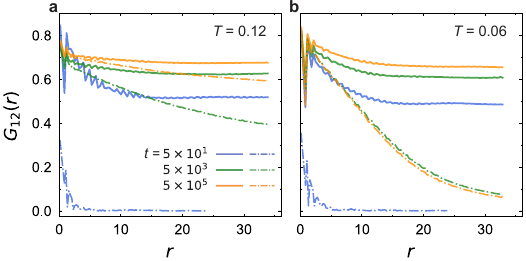}
    \caption{{\bf Comparison of DDQC growth in systems starting from the mechanical approach and the quenched liquid.} {\bf a,b} Time evolution of the correlation function of the $12$-fold order parameter, $G_{12}(r)$, at $T=0.12$ and $0.06$, respectively. Solid and dot-dashed lines are for states from the mechanical approach and liquid state, respectively. The legend indicates the specific instants used for comparison. The same TLS systems as in Fig.~\ref{fig:fig4} are used.}
    \label{fig:fig5}
\end{figure}

In Fig.~\ref{fig:fig5}a, we compare the time evolution of the correlation function of the 12-fold bond orientational order, $G_{12}(r)$ (detailed in Methods), at $T=0.12$ for two initial conditions. Starting from a state generated by the mechanical approach, $G_{12}(r)$ exhibits a plateau at large distances $r$, indicating the presence of a long-range DDQC order. The plateau value increases over time, demonstrating the growth of the overall DDQC order. In contrast, starting from a liquid state quenched from a high temperature, $G_{12}(r)$ is small at short times. At longer times, $G_{12}(r)$ decays with increasing $r$, suggesting the absence of a long-range order. With the increase of time, $G_{12}(r)$ increases and becomes more flattened, indicating the growth of the DDQC order. However, it remains smaller than that from the mechanical approach. This comparison highlights the efficiency of the mechanical approach in facilitating the formation of DDQCs. For the same amount of time, the mechanical approach enables significantly faster formation of DDQCs with a higher quality.

Figure~\ref{fig:fig5}b further showcases the effectiveness of the mechanical approach. At a lower temperature ($T=0.06$), the growth of the DDQC order from the liquid state is significantly hindered within the same time scale as in Fig.~\ref{fig:fig5}a. $G_{12}(r)$ evolves rather slowly and exhibits a rapid decay with increasing $r$ even at long times. In contrast, the time evolution of the state from the mechanical approach remains comparable to that at $T=0.12$. Therefore, at low temperatures when the self-assembly of DDQCs from a liquid state is impeded by the sluggish particle dynamics, the mechanical approach can still promote structural relaxation and the growth of the DDQC order in a non-diffusive manner. In addition to facilitating the DDQC formation, the mechanical approach reveals a distinct pathway for the DDQC growth.   

\begin{figure}
    \centering
    \includegraphics[width=0.95\linewidth]{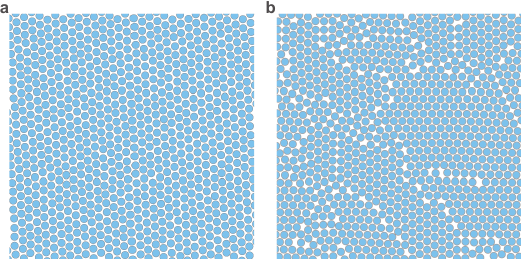}
    \caption{{\bf Example of the failure of the mechanical approach to induce quasicrystalline order in a non-DDQC-forming system.} {\bf a} A state of TLS disks after thermal equilibration of an $N=900$ liquid at $T=0.12$ and $\rho=1.02$. The equilibrium state is the hexagonal phase. {\bf b} A packing of the same system in {\bf a} generated by introducing $n_{\rm v}=70$ vacancy pairs to an $N=1089$ square packing and minimizing the energy. The mechanical approach results in a polycrystalline state of hexagonal phase rather than DDQC structures. }
    \label{fig:fig6}
\end{figure}

\vspace{2mm}
\noindent {\bf \large{Discussion}}
\vspace{1mm}

\noindent By perturbing the square packing of hard disks, we realize the spontaneous formation of a DDQC approximant and the local formation of DDQC motifs via a purely mechanical approach. We further demonstrate that this mechanical approach can facilitate the self-assembly of DDQCs in known DDQC-forming systems. Our approach shows high efficiency in forming a robust DDQC order athermally, accelerating the DDQC growth with the aid of thermal treatment, and enabling the DDQC growth even at rather low temperatures when self-assembly of DDQCs from liquid states is significantly suppressed. 

To our knowledge, the spontaneous formation of complex crystals and DDQC structures has not been anticipated previously in the simplest system of identical hard disks. Our findings thus challenge established theories and expand our understanding of DDQCs and their approximants. Moreover, the non-diffusive growth of the DDQC order triggered by our mechanical approach provides insights into the growth pathways of DDQCs. Our results can be validated in experimental systems, such as colloidal suspensions using optical tweezers to construct and manipulate the square packing. 

Furthermore, our study proposes a method for predicting whether a system can self-assemble into a DDQC under specific conditions. As illustrated in Fig.~\ref{fig:fig6}, at densities where DDQCs are thermodynamically unstable, our mechanical approach would not yield any DDQC structures. Therefore, the emergence of DDQC structures via our mechanical approach may serve as an indicator of the DDQC-forming capability. 

While our study focuses on DDQCs, it would be interesting to investigate whether quasicrystals with other symmetries can be generated using our mechanical approach. We expect that our approach may be effective for quasicrystals with squares as a tiling element, such as octagonal quasicrystals. However, for quasicrystals lacking squares as a tiling element, it remains unclear whether the mechanical approach would be effective with the design of alternative unstable initial structures. These questions warrant further exploration in follow-up studies.   

\vspace{2mm}
\noindent {\large{\bf Methods}}
\vspace{1mm}

\noindent {\bf Models}

\noindent Our systems are square cells with sidelength $L$, consisting of $N$ identical particles (disks) with mass $M$. Periodic boundary conditions are applied in both directions. We study two types of particle interactions to simulate behaviors of hard disks and a DDQC-forming system, respectively. 

The first interaction is the harmonic repulsion. The interaction potential between particles $i$ and $j$ is
\begin{equation}\label{pot:harmonic}
U\left( r_{ij}\right)  
= \frac{\epsilon}{2} \left( 1-\frac{r_{ij}}{\sigma} \right) ^2 \Theta\left( 1-\frac{r_{ij}}{\sigma}\right), 
\end{equation}
where $\epsilon$ is the characteristic energy scale of the interaction, $r_{ij}$ is the separation between the two particles, $\sigma$ is the disk diameter, and $\Theta(x)$ is the Heaviside step function. In the zero temperature and pressure limits ($T\to 0$ and $p\to 0$), particles interacting via the harmonic repulsion behave like hard particles \cite{Xu_Haxton_Liu_Nagel_2009}.  

The second interaction potential is designed with two length scales to stabilize DDQCs \cite{Kryuchkov_Yurchenko_Fomin_Tsiok_Ryzhov_2018,Padilla_Ramírez-Hernández_2020,Coli_Boattini_Filion_Dijkstra_2022}:
\begin{equation}\label{pot:coresoftened}
U\left(r_{ij}\right) = \epsilon
\left( \frac{\sigma}{r_{ij}} \right) ^{14} + \frac{\epsilon}{2} \left[ 1-\tanh \left( k r_{ij}-k\sigma_1\right) \right], 
\end{equation}
which is the inverse-power-law potential followed by a shoulder, where $\sigma$ and $\sigma_1$ are characteristic length scales of the core and the shoulder, respectively, and $k$ sets the relative height of the shoulder. Under appropriate conditions, particles interacting via this TLS potential can self-assemble into DDQCs in thermodynamic equilibrium ~\cite{Kryuchkov_Yurchenko_Fomin_Tsiok_Ryzhov_2018}. We use $\sigma_1=1.36\sigma$ and $k=10.0\sigma^{-1}$, slightly different from Ref.~\cite{Kryuchkov_Yurchenko_Fomin_Tsiok_Ryzhov_2018} to achieve better DDQC order. 

For both potentials, we set units of energy, length, and mass to be $\epsilon$, $\sigma$, and $M$. The time is thus in units of $M^{1/2} \sigma \epsilon^{-1/2}$. The temperature is in units of $\epsilon k_{\rm B}^{-1}$ with $k_{\rm B}$ being the Boltzmann constant. For the harmonic potential, we tessellate the packings into polygons by connecting particles in contact with bonds. For the TLS potential, we connect particles whose separation is smaller than $\sigma_1$ with bonds \cite{Kryuchkov_Yurchenko_Fomin_Tsiok_Ryzhov_2018,Padilla_Ramírez-Hernández_2020}.

\vspace{2mm}
\noindent {\bf Simulation methods}

\noindent For systems with the harmonic potential, we obtain static particle packings at a given pressure by minimizing the enthalpy $H=U+pL^2$ via the fast inertial relaxation engine algorithm~\cite{Bitzek_Koskinen_Gähler_Moseler_Gumbsch_2006}, where $U$ is the total potential energy summed over all pairs of interacting particles. We use a constant pressure of $p=10^{-4}$. 

For systems with the TLS potential, we minimize the total potential energy $U$ at a given number density $\rho=NL^{-2}$ to obtain static packings. We also conduct molecular dynamic simulations in the canonical ensemble using lammps~\cite{LAMMPS} to obtain the spatiotemporal evolution of the system. When the number density $\rho$ is around $0.94$, the system self-assembles into a DDQC at low temperatures. 

\vspace{2mm}
\noindent {\bf Diffraction pattern}

\noindent The diffraction pattern is characterized by the static structure factor: \(S(\mathbf{q}) = \frac{1}{N} \left< \rho(\mathbf{q}) \rho(-\mathbf{q}) \right>,\) where $\rho(\mathbf{q}) = \sum_{j=1}^N {e^{{\rm i}\mathbf{q}\cdot\mathbf{r}_j}}$ is the Fourier transform of the density, $\mathbf{r}_j$ is the position of particle $j$,  $\mathbf{q}$ is the wave vector satisfing periodic boundary conditions, and $\left\langle . \right\rangle$ denotes the ensemble average. 

\vspace{2mm}
\noindent {\bf Order parameter and correlation function}

\noindent For a particle at $\mathbf{r}_i$, we define its $m$-fold bond orientational order parameter as $\varPsi_{m}\left( \mathbf{r}_i\right)= \frac{1}{n_{\rm b}}\sum_{l=1}^{n_{\rm b}}e^{{\rm i}m\theta\left(\mathbf{r}_i-\mathbf{r}_l\right)}$, where the sum is over all $ n_{\rm b}$ neighbors, and $\theta\left(\mathbf{r}_i-\mathbf{r}_l\right)$ is the angle between $\mathbf{r}_i-\mathbf{r}_l$ and a reference direction. The average bond orientation order is defined as $\chi_m = \left< |\varPsi_m(\mathbf{r}_i)|^2\right>$, where $\left\langle . \right\rangle$ denotes the average over particles and configurations \cite{Dotera_Oshiro_Ziherl_2014}. Here we mainly show results for $m=4, 6$, and $12$.

The correlation function of the $m$-fold bond orientational order is defined as $G_{m}(r)= \left\langle \varPsi_{m}^*\left( \mathbf{r}_i\right) 
\varPsi_{m}\left( \mathbf{r}_j\right) \right\rangle, $ where $r=|\mathbf{r}_i - \mathbf{r}_j|$ and $\left\langle . \right\rangle$ denotes the average over all pairs of particles and configurations. For DDQCs, $G_{12}(r)$ is used to characterize the quasicrystalline order \cite{Dotera_Oshiro_Ziherl_2014}. 

\vspace{2mm}
\noindent {\bf \large{Data availability}}

\noindent The data that support the findings of this study are included in the article and/or the Supporting Information and are available from the corresponding authors upon request.

\vspace{2mm}
\noindent {\bf \large{Code availability}}

\noindent The computer codes of this study are available from the corresponding authors upon request.

\vspace{2mm}
\noindent {\bf Acknowledgements}

\noindent We thank Peng Tan for useful discussions. We acknowledge the support from the National Natural Science Foundation
of China (Grant Nos. 12334009 and 12274392). 

\vspace{2mm}
\noindent {\bf Author contributions}

\noindent N.X. designed the project. Z.J. performed the simulations. Z.J., J.Z., M.Z., J.S., J.H., H.T., and N.X. analyzed the data, Z.J., H.T., and N.X. wrote the paper. H.T. and
N.X. supervised the project. 

\vspace{2mm}
\noindent {\bf Competing interests}

\noindent The authors declare no competing interests.

\vspace{2mm}
\noindent {\bf Correspondence} and requests for materials should be addressed to Hua Tong or Ning Xu.

\clearpage
\newpage
\renewcommand{\thefigure}{\arabic{figure}}
\renewcommand{\figurename}{{\bf Extended Data Fig.}}

\setcounter{figure}{0}

\end{document}